\documentclass[aps,pra,twocolumn,groupedaddress,floatfix,superscriptaddress]{revtex4-1}

\usepackage{amsmath}
\usepackage{amssymb}
\usepackage{graphicx}
\usepackage{float}
\usepackage[english]{babel}
\usepackage{dsfont}
\usepackage{epstopdf}
\usepackage{soul}
\usepackage{color}
\usepackage{braket}

\usepackage[colorlinks=true,
            linkcolor=magenta,
            urlcolor=blue,
            citecolor=blue]{hyperref}

\usepackage[normalem]{ulem}

\def\be{\begin{equation}}
\def\ee{\end{equation}} 
\def\bea{\begin{eqnarray}}
\def\eea{\end{eqnarray}} 
\def\ba{\begin{array}} 
\def\ea{\end{array}}
\def\pa{\partial}

\def\om{\omega}

\def\nn{\nonumber}
\def\ket{\rangle}

\def\f{\frac}

\def\ra{\rightarrow}

\newcommand{\mv}[1]{\langle #1\rangle}

\emergencystretch=\maxdimen
\begin{document}
\title{Nonlinear dynamics of Aharonov-Bohm cages}
\author{Marco Di Liberto}
\email{mar.diliberto@gmail.com}
\affiliation{Center for Nonlinear Phenomena and Complex Systems, Universit\'e Libre de Bruxelles, CP 231, Campus Plaine, B-1050 Brussels, Belgium}

\author{Sebabrata~Mukherjee}
\affiliation{Scottish Universities Physics Alliance (SUPA), Institute of Photonics and Quantum Sciences, School of Engineering and Physical Sciences, Heriot-Watt University, Edinburgh EH14 4AS, United Kingdom}
\affiliation{Department of Physics, The Pennsylvania State University, University Park, PA 16802, USA}

\author{Nathan~Goldman}
\affiliation{Center for Nonlinear Phenomena and Complex Systems, Universit\'e Libre de Bruxelles, CP 231, Campus Plaine, B-1050 Brussels, Belgium}

\begin{abstract}

The interplay of $\pi$-flux and lattice geometry can yield full localization of quantum dynamics in lattice systems, a striking interference phenomenon known as Aharonov-Bohm caging. At the level of the single-particle energy spectrum, this full-localization effect is attributed to the collapse of Bloch bands into a set of perfectly flat (dispersionless) bands. In such lattice models, the effects of inter-particle interactions generally lead to a breaking of the cages, and hence, to the spreading of the wavefunction over the lattice. Motivated by recent experimental realizations of analog Aharonov-Bohm cages for light, using coupled-waveguide arrays, we hereby demonstrate that caging always occurs in the presence of local nonlinearities. As a central result, we focus on special caged solutions, which are accompanied by a breathing motion of the field intensity, that we describe in terms of an effective two-mode model reminiscent of a bosonic Josephson junction. 
Moreover, we explore the quantum regime using small particle ensembles, and we observe quasi-caged collapse-revival dynamics with negligible leakage. 
The results stemming from this work open an interesting route towards the characterization of nonlinear dynamics in interacting flat band systems.

\end{abstract}

\maketitle

{\it Introduction.$-$} The realization of synthetic gauge fields in quantum-engineered matter~\cite{cooper2008rapidly,dalibard2011colloquium,Goldman2014} and photonics~\cite{Lu2014,Ozawa2018} has revolutionized the realm of quantum simulation, by offering the possibility of studying exotic states of matter in a well-controlled environment. In lattice systems, this has led to the exploration of topological phenomena reminiscent of the quantum Hall effects and topological insulators~\cite{goldman2016topological,Ozawa2018, Cooper2018}, and to the study of frustrated magnetism~\cite{struck2011quantum,struck2013engineering,eckardt2017colloquium}. 

While arbitrary synthetic magnetic fluxes can be realized in artificial lattices~\cite{aidelsburger2018artificial}, such as optical lattices for ultracold gases or photonics lattices for light, combining these artificial fields with strong inter-particle interactions still remains a fundamental challenge. An appealing strategy by which interaction effects can be enhanced in lattice systems consists in designing models that exhibit flat (dispersionless) Bloch bands. In these situations, interactions indeed set the dominant energy scale, hence potentially leading to intriguing strongly-correlated phenomena~\cite{Huber2010,Bernevig2011,Peotta2015}, examples of which include the celebrated fractional quantum Hall effect~\cite{Stormer1982,yoshioka2013quantum} and the recently discovered high-$T_c$ superconductivity in twisted bilayer graphene~\cite{Cao2018,cao2018correlated}. This highly motivates the implementation of a wide range of flat-band models in artificial lattice systems~\cite{Mukherjee2015observation,Vicencio2015observation,Takahashi2015,Leykam2018a}.

Interestingly, specific lattice models exhibit a striking interference phenomenon, called Aharonov-Bohm (AB) caging, by which the single-particle spectrum collapses into a set of perfectly flat (dispersionless) Bloch bands~\cite{Vidal1998aharonov}. This remarkable effect can be found in the one-dimensional rhombic lattice~\cite{Mukherjee2015rhombic} or in the two-dimensional dice lattice~\cite{bercioux2009massless,bercioux2011topology}, in the presence of a magnetic $\pi$-flux (i.e.~half a flux quantum) per plaquette~\cite{Vidal1998aharonov}, and it can be attributed to destructive interferences that fully localize any initially prepared wavefunction. AB cages were first observed in networks of conducting wires~\cite{Abilio1999magnetic,naud2001aharonov}, and were recently realized in photonic lattices~\cite{Mukherjee2018,Szameit2018}.

The impact of interactions on AB cages has been investigated in different regimes. At the two-body level, the characteristic full-localization property of AB cages was shown to be substantially altered by interactions, which couple distinct single-particle states localized at adjacent unit cells, and hence introduce a mechanism by which the two-particle wavefunction can spread over the entire lattice~\cite{Vidal2000interaction, Creffield2010}. At the many-body level, the new channels offered by the interactions are responsible for the appearance of supersolidity at low filling~\cite{Moller2012correlated}, or pair condensation at commensurate filling~\cite{Doucot2002pairing,Rizzi2018}; see also Refs.~\cite{Aoki2013,Huber2013,Huber2018}. However, not much is known regarding the fate of AB cages in the presence of mean-field interactions, as captured by the Gross-Pitaevskii equation. This scenario is particularly relevant to photonic systems, \emph{e.g.}~arrays of coupled optical waveguides or exciton-polariton pillars~\cite{Carusotto2013}, where the non-linearity of the medium becomes significant at sufficiently high field intensity. 

Characterizing the robustness of AB caging in the presence of mean-field interactions is particularly motivated by their recent realization in photonics~\cite{Mukherjee2018,Szameit2018}, where the propagation of intense light waves inside coupled-waveguide arrays can be captured by a discrete nonlinear Schr\"odinger (Gross-Pitaevskii-type) equation~\cite{eisenberg1998discrete, szameit2010discrete, Lederer2008}. 
We point out that the interplay of AB caging and strong (beyond mean-field) interactions  could be explored in superconducting circuits~\cite{Simon2018} or in ultracold atoms in optical lattices~\cite{bloch2012quantum}. 

In this paper, we investigate the fate of AB caging in the presence of local mean-field  interactions and discuss the optical-waveguides platform of Ref.~\cite{Mukherjee2018} as a relevant experimental platform. We demonstrate the survival of AB caging and focus on the dynamics of specific initial states to highlight a characteristic breathing motion of the field intensity, which we describe using a simple effective two-mode model that is directly analogous to a bosonic Josephson junction. This analogy offers an intriguing link with the physics of weakly-interacting Bose-Einstein condensates in tilted double wells~\cite{Oberthaler2007}. We finally go beyond the mean-field framework by studying the caging of small particle ensembles deep in the quantum regime.

\begin{figure}[!t]
\center
\includegraphics[width=0.9\columnwidth]{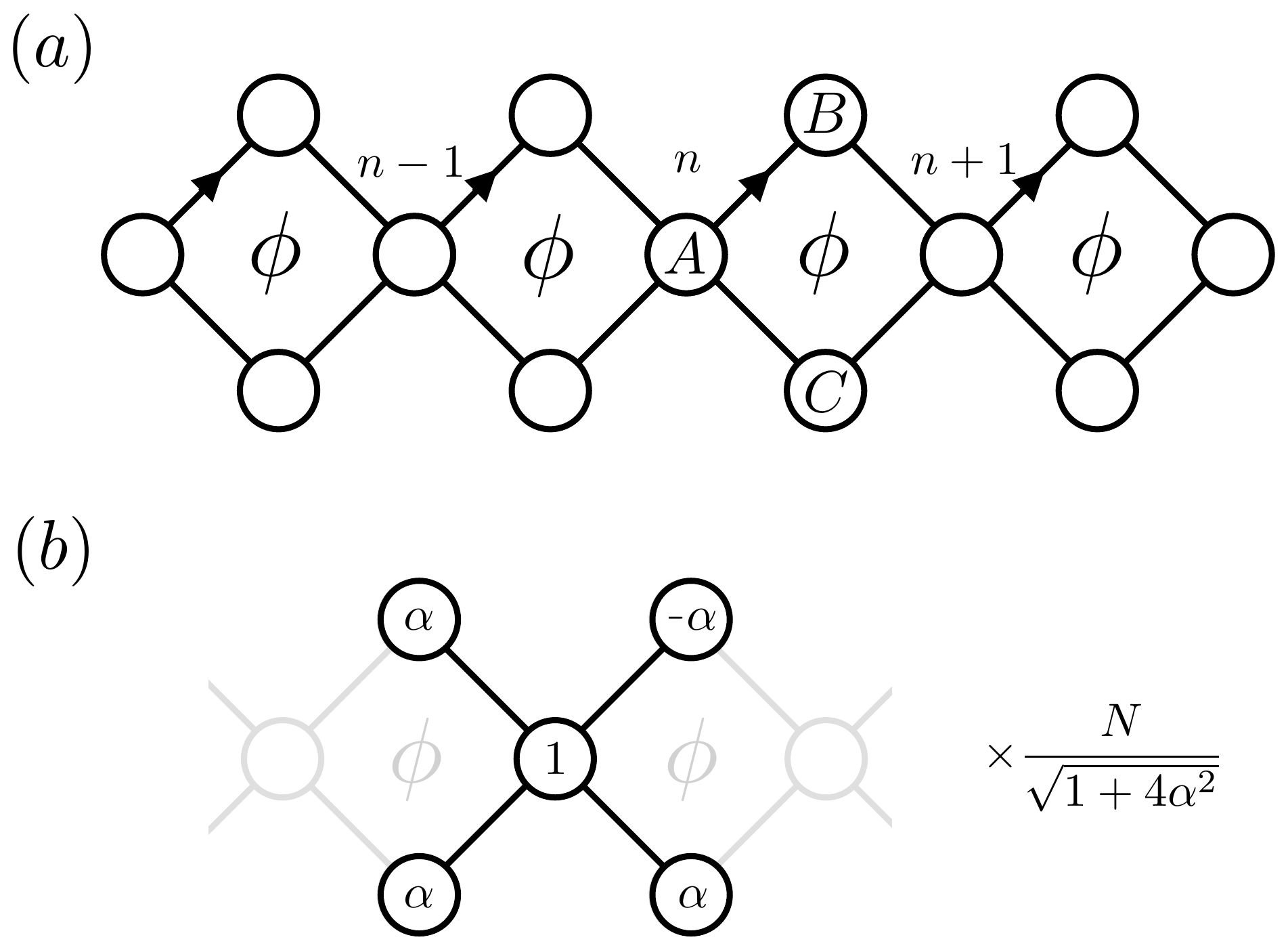}

\caption{(a) Rhombic lattice with flux $\phi$ per plaquette. The gauge choice for the Peierls phase factors in Eqs.~\eqref{eq:GP_rhombic} is represented by arrows (see intra-cell $A\!-\!B$ bonds). (b) Configuration used as the input for nonlinear caging dynamics; here $0\leq \alpha \leq 1$. In the non-interacting limit, this configuration with $\alpha\!=\!\pm 1/2$ corresponds to states in the upper and lower bands, respectively. The normalization constant, which depends on the total number of particles $N$, is shown.
}
\label{fig:lattice}
\end{figure}


{\it Model.$-$} We consider the dynamics of a classical field on a one-dimensional rhombic (diamond) chain with non-vanishing flux $\phi$ per plaquette and onsite nonlinearity (interaction) $U$, as described by the nonlinear Schr\"odinger equations
\begin{align}
\label{eq:GP_rhombic}
i\, \pa_t a_n &= - J (b_{n-1} + c_{n-1} + e^{i\phi} b_n + c_n )+ U | a_n |^2 a_n \,, \nn\\
i\, \pa_t b_n &= - J (e^{-i\phi} a_{n-1} + a_{n} )+ U | b_n |^2 b_n \,, \nn\\
i\, \pa_t c_n &= - J (a_{n-1} + a_{n} )+ U | c_n |^2 c_n \,.
\end{align}
The quantities $a_n(t)$, $b_n(t)$ and $c_n(t)$ are the field amplitudes for the sites $A_n$, $B_n$ and $C_n$ in the $n$th unit cell; see Fig.~\ref{fig:lattice}(a) for a definition of the sites indices. The parameter $J$ denotes the hopping amplitude between neighboring sites. For convenience, a gauge choice is made in Eq.~\eqref{eq:GP_rhombic} such that the flux within each plaquette $\phi$ enters the model through a single Peierls phase factor~\cite{Hofstadter_ref} in each unit cell. 

For a flux $\phi \!=\! \pi$ and negligible nonlinearity ($U\!=\!0$), the system displays three flat bands at energies $\epsilon_\pm \!=\! \pm2J$ and $\epsilon_0\!=\!0$, a regime known as Aharonov-Bohm caging: the spectrum only displays fully localized eigenstates~\cite{Vidal1998aharonov}. The corresponding eigenstates can be written as $|v_\pm\ket = |B_{n-1}\ket + |C_{n-1}\ket \mp 2 |A_{n}\ket - |B_{n}\ket + |C_{n}\ket$ and $|v_0\ket = |B_{n-1}\ket + |C_{n-1}\ket + |B_{n}\ket - |C_{n}\ket$, respectively. When $\phi\neq\pi$ the zero-energy flat band survives, but the upper and lower bands become dispersive.


{\it Caging solutions in the presence of nonlinearities.$-$} We henceforth focus on the caging limit $\phi\!=\!\pi$. The full localization of the spectrum is expected to break in the presence of interactions, since the latter couple distinct neighboring localized states, thus introducing a mechanism for the spreading of the wavefunction across the lattice; see, for instance, Ref.~\cite{Vidal2000interaction} for the two-body case, and Ref.~\cite{Rizzi2018} for the many-body case and a discussion on the interaction terms arising among flat band states. However, as it can be proven from Eqs.~\eqref{eq:GP_rhombic}, caging still occurs in the presence of mean-field local interactions and the corresponding dynamics is robust with respect to delocalization for any arbitrary initial state. A pivotal role in this mechanism is played by the hub sites $A_n$, which can block the spreading over time even in the presence of nonlinearities. The resulting dynamics thus remains confined (caged) between two $A$ sites. Locality is a crucial ingredient in this case, since long-range nonlinear terms would instead break the caging. We have validated the previous statements by using random initial configurations:~while caged dynamics is always observed in the presence of local or short-ranged (nearest-neighbor) interactions, they are already substantially deteriorated in the case of next-nearest-neighbor interactions~\cite{SuppMat}. 

For the sake of simplicity, we restrict ourselves to caged solutions that are characterized by vanishing amplitudes at the lattice sites $A_{n-1}$ and $A_{n+1}$ at all times, thus leaving the dynamics confined in the intermediate five sites; see Fig.~\ref{fig:lattice}(b). Setting $a_{n-1}(t)\!=\!a_{n+1}(t)\!=\!0$ together with $\dot a_{n-1}(t)\!=\!\dot a_{n+1}(t)\!=\!0$ yields the conditions $b_{n-1}(t)\!=\!c_{n-1}(t)$ and $b_n(t)\!=\!-c_n(t)$. We decompose the three independent complex numbers $b_{n-1}(t)$, $a_{n}(t)$ and $b_{n}(t)$ into amplitude and phase, and we consider the case of a symmetric time-evolution $|b_{n-1}(t)| \!=\! |b_n(t)|$. After some algebra, one finds that this ansatz for the dynamics fixes a condition for the phases, namely $\mathrm{arg}[b_n(t)/b_{n-1}(t)] \!=\! \pi$ unless $|a_n(t)|\!=\!0$. We therefore conclude that states of the form shown in Fig.~\ref{fig:lattice}(b) are caged solutions to the nonlinear equations of motion~\eqref{eq:GP_rhombic}. We point out that this specific field configuration actually shares the same phase profile as that of the lowest-energy eigenstates $|v_- \ket$ associated with the single-particle spectrum. 

\begin{figure}[!t]
\center
\includegraphics[width=\columnwidth]{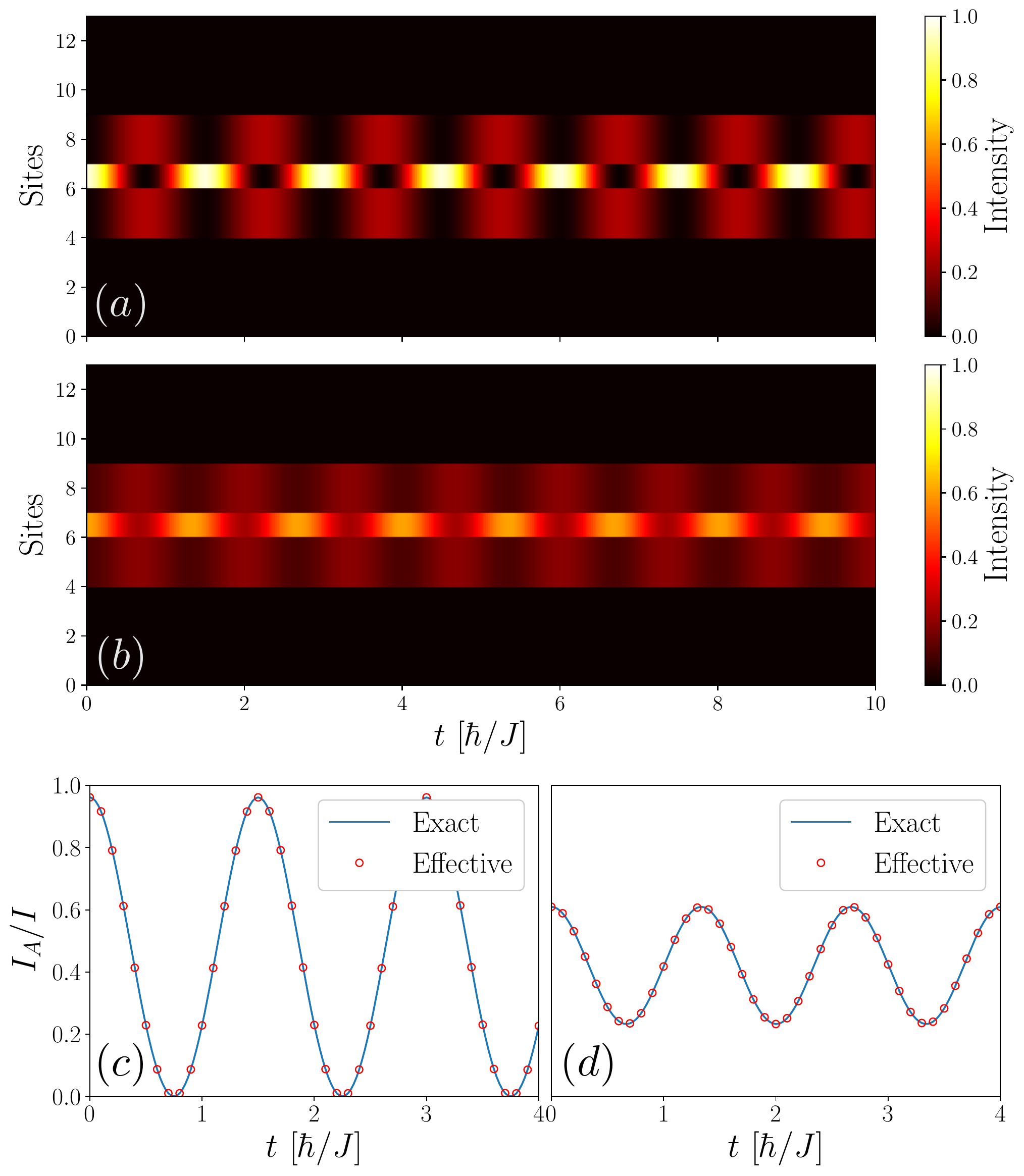}
\caption{(a-b) Time-evolution of the intensity (particle density), $I\!=\!\vert \psi_n \vert^2$, on a diamond chain with $\pi$-flux and $L\!=\!13$ sites, for $g\!=\!2.5J$: (a) $\alpha\!=\!0.1$ and (b) $\alpha\!=\!0.4$. Here, the density is normalized with respect to the total number of particles $N$. A breathing motion together with the caging of the total intensity is observed. (c-d) Time-evolution of the intensity $I_A$, associated with the central $A$ site in Fig.~\ref{fig:lattice}(b), as obtained from the exact Gross-Pitaevskii evolution (line) and from the effective two-mode model (circles). All parameters are the same as in (a)-(b).}
\label{fig:map_dyn}
\end{figure}

Consequently, we are left with only two independent quantities, which we parametrize as 
\begin{align}
\label{eq:def_dw}
a_n(t) &\equiv \sqrt{\f{N-n(t)}{2}}\, e^{i\theta(t)}\,, \nn \\
b_{n-1}(t) &\equiv \sqrt{\f{N+n(t)}{8}}\, e^{i\varphi(t)}\,,
\end{align}
where we have defined the conserved total number of particles in the cage $N=|a_n|^2 +  4|b_{n-1}|^2$.  The relevant degrees of freedom are the fractional particle imbalance $z(t) \equiv n(t)/N$, with $-1\leq\!z\!\leq 1$, and the phase difference $\xi(t) \equiv \theta(t) - \varphi(t)$, which satisfy the following coupled nonlinear equations,
\begin{align}
\label{eq:eff_model}
\dot z(t) &= - 4J \sqrt{1-z(t)^2} \sin \xi(t) \,, \nn\\
\dot \xi(t) &= 4J \f{z(t)}{\sqrt{1-z(t)^2}} \cos \xi(t) + \f 5 8 g z(t) - \f 3 8 g\,,
\end{align}
where we have defined the nonlinear coupling $g\equiv N U$. 

Interestingly, the coupled equations \eqref{eq:eff_model} can be obtained from the classical Hamiltonian
\be
\label{eq:ham_eff}
\mathcal H = -4J\sqrt{1-z^2}\cos\xi + \f{5}{16} g z^2 - \f 3 8 g z\,,
\ee
by considering $\xi$ as a generalized coordinate and $z$ as its canonically conjugate momentum. In this respect, Eqs.~\eqref{eq:eff_model}-\eqref{eq:ham_eff} describe the dynamics of a non-rigid pendulum, and $\mathcal H$ describes the conserved energy of the system. This shows how restricting the dynamics to the solutions in Eq.~\eqref{eq:def_dw}, whose equations of motion correspond to Eqs.~\eqref{eq:eff_model}, allows one to map the initial problem onto a two-mode model for an amplitude degree of freedom and a phase degree of freedom. This mapping offers an intriguing reinterpretation of the nonlinear AB-caging dynamics in terms of that associated with a weakly-interacting Bose condensate in a tilted double well~\cite{Smerzi1997,Smerzi1999} or in two hyperfine states~\cite{Oberthaler2015}. In other words, this non-trivial dynamics corresponds to that of a generalized bosonic Josephson junction~\cite{Cataliotti2001,Oberthaler2007}, which displays, for instance, macroscopic self-trapping for high-energetic excitations~\cite{Oberthaler2005} and ``twist-and-turn" spin squeezing~\cite{Oberthaler2015}. 

Moreover, the ground state of $\mathcal{H}$, occurring for $\xi=0$ and for values of $z_0$ that depend on the ratio $g/J$, corresponds to a stationary discrete soliton solution that can be continuously connected to the single particle eigenstates when $g/J \ra 0$. These solutions typically occur in flat-band systems, occupy few sites and have no exponential tail \cite{Lederer2008, Flach2008, Flach2018}.


{\it Time-evolution of nonlinear caged states.$-$} We now focus on the dynamics in the presence of $\pi$-flux, and we reveal a nonlinear caging characterized by a breathing motion of the field intensity inside the cage. 

Based on the results presented in the previous paragraph, we consider an initial state at $t\!=\!0$ of the form $b_{n-1}(0) \!=\! c_{n-1}(0) \!=\! - b_n(0) = c_n(0)$ and $a_n(0) \neq 0$, and we define the parameter $\alpha \!=\! b_{n-1}(0)/a_{n}(0)$. Furthermore, we fix the initial phase difference to $\xi(0) = 0$, see Fig.~\ref{fig:lattice}(b), which amounts to take a real valued $\alpha\!>\!0$. Together with the interaction parameter, $g\!=\!NU$, these parameters uniquely determine the initial conditions of the problem. For simplicity, we will consider units where $U\!=\!J\!=\!1$ in the rest of the discussion, which potentially corresponds to a regime of large nonlinearities.

\begin{figure}[!t]
\center
\includegraphics[width=\columnwidth]{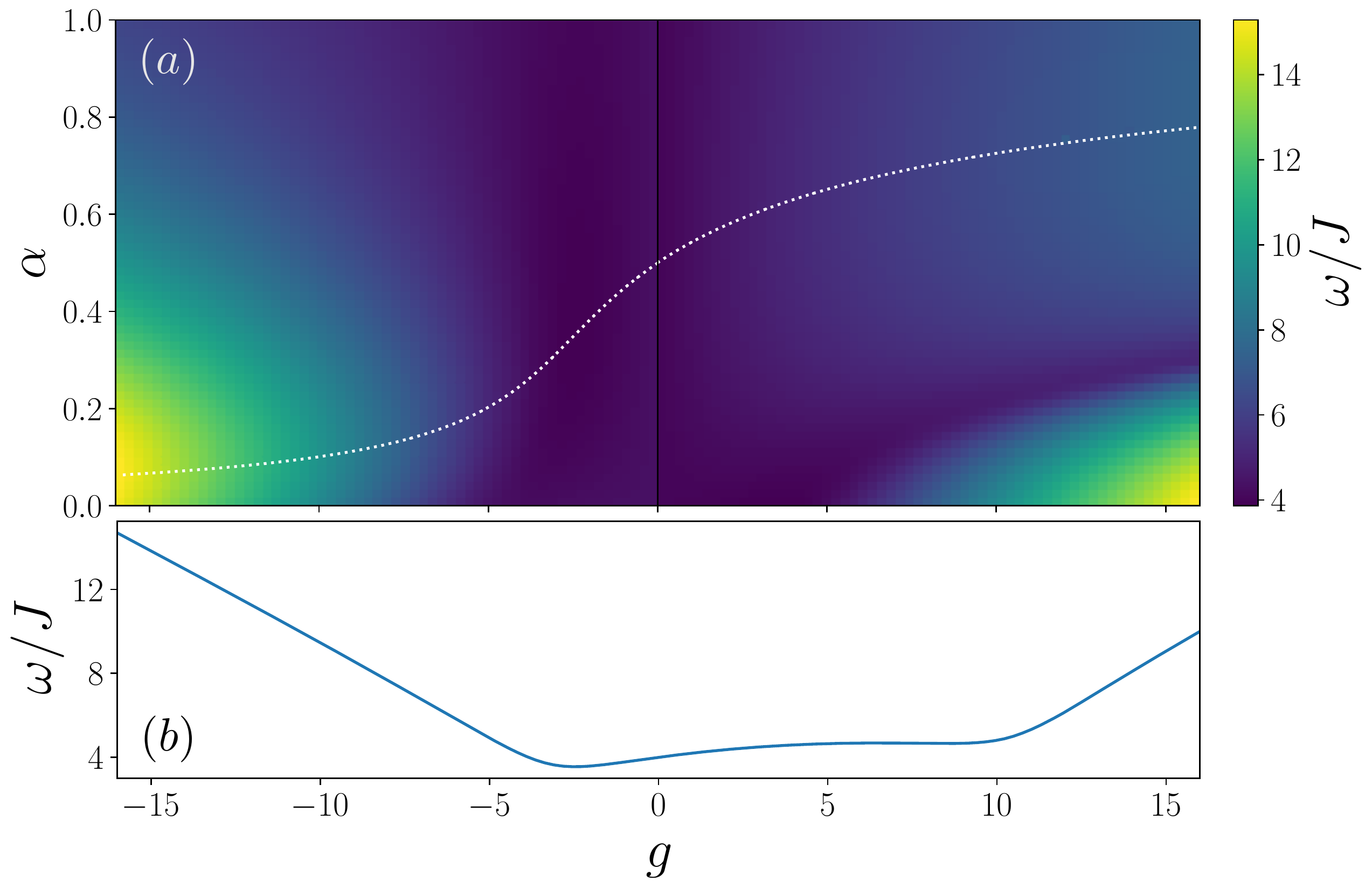}
\caption{(a) Main frequency of the breathing dynamics for (attractive) $g<0$ and (repulsive) $g>0$ interactions. The white dashed line corresponds to the minimum of the classical Hamiltonian $\mathcal H$ in Eq.~\eqref{eq:ham_eff} occurring for $\xi=0$, namely a stationary discrete soliton solution of the Gross-Pitaevskii equations. When $g\ra0$, the minimum reaches $\alpha \ra 1/2$, which corresponds to the localized states of the single-particle lowest energy band. (b) Cut in (a) at $\alpha=0.2$.
}
\label{fig:map_period}
\end{figure}

In Fig.~\ref{fig:map_dyn}(a-b), we show the resulting periodic dynamics, for a nonlinear regime corresponding to $g\!=\!2.5 J$, and for two different initial conditions ($\alpha\!=\!0.1$ and $\alpha\!=\!0.4$). In Fig.~\ref{fig:map_dyn}(c-d), we demonstrate that the full Gross-Pitaevskii dynamics can be quantitatively described by the two-mode dynamical equations~\eqref{eq:eff_model}. The dynamics that we have presented so far corresponds to the case where one prepares an initial state that is characterized by a non-vanishing displacement $z(0)\!\neq\!0$ with respect to the minimum (ground state) of the classical Hamiltonian $\mathcal H$ in Eq.~\eqref{eq:ham_eff}; see discussion above on the stationary solutions of the equations of motion and the white dashed line in Fig.~\ref{fig:map_period}(a). When $z(0)\!\approx\!z_0$, the system is in the regime of small oscillations and one expects purely harmonic dynamics, whereas the dynamics becomes anharmonic when $z(0)$ is sufficiently far from $z_0$ (as revealed, for instance, by the Fourier spectrum). 

We have investigated caging in a large region of parameter space, and we characterize the resulting periodic breathing dynamics by representing its main (smallest) frequency $\omega$ in Fig.~\ref{fig:map_period}, both for $g\!>\!0$ and $g\!<\!0$. The main frequency of these oscillations displays a non-monotonic behavior as a function of $g$, and it is given by $\om\!=\!4J$ when $g\ra 0$, as expected from the fact that the initial state overlaps with the upper and lower flat bands of the single-particle spectrum in this limit. We note that the periodicity of the oscillations at small negative $g$ is in agreement with the results of Ref.~\cite{Gligoric2018}.

{\it Experimental considerations and beyond mean-field.$-$}  We have shown that AB caging survives in a rhombic lattice with $\pi$-flux in the presence of mean-field local interactions independently of the strength of nonlinearities.  Similarly to the single-particle case, caging is a consequence of the existence of hub sites $A_n$, see Fig.~\ref{fig:lattice}(a), which block the spreading of the wavefunction thanks to phase interferences. Here, locality is essential and we have observed breaking of caging only when long-range interactions (\emph{i.e.}~at least next-nearest-neighbor) are present~\cite{SuppMat}. For onsite interactions and when the cage includes only one hub site $A$, the corresponding breathing motion of the density can be understood in terms of a simple two-mode theory with a period of oscillation that displays a non-monotonic behavior. 

Imperfections (\emph{e.g.} the presence of disorder \cite{Vidal2001,Molina2018}, deviations from the initial conditions or from the $\pi$-flux limit) can potentially lead to instabilities in the dynamics, which in turn will alter the interference processes and the resulting time evolution. However, if the instabilities associated with these modes are weak, namely, if the instability manifests itself within a time-scale $\tau$ such that $\omega \tau\!>\!1$, where $\omega$ is the frequency of the expected oscillations, one should still be able to observe the nonlinear dynamics associated with the specific (ideal) initial conditions introduced in this work. A linear stability analysis performed on the state represented in Fig.~\ref{fig:lattice}(b) would allow one to extract the relevant time scale $\tau$, and thus identify the optimal set of parameters for a given experimental setting; see also the discussion on stability in the numerical study reported in Ref.~\cite{Gligoric2018}.
 
Ultrafast-laser-fabricated waveguide arrays are a promising platform to observe nonlinear caging dynamics, especially in light of the recent experimental realizations~\cite{Mukherjee2018,Szameit2018}. In this platform, the propagation of the electric field is described by an (analog) discrete Gross-Pitaevskii equation, where the mean-field interaction is described by the optical Kerr nonlinearity of the medium~\cite{szameit2010discrete, Carusotto2013, Ozawa2018} and the waveguide-propagation coordinate plays the role of time. The largest ``interaction" strength $g$ that can be achieved in coupled-waveguide arrays depends on the nonlinear refractive index of the medium, on the effective area of the waveguide mode, and on the wavelength and the power of the incident light~\cite{eisenberg1998discrete}. In order to provide an accurate estimation of realistic nonlinearities, we have performed a preliminary experimental analysis of these effects in a coupled-waveguide array realized in a borosilicate glass substrate. Our measurements indicate that one can reach values of the order of $|g/J|\!\approx\!10$, thus making the exploration of our results possible in current experiments. Besides, we note that the short-time $(t\!\sim\!4 \hbar/J)$ dynamics is relatively easy to access experimentally, which should allow one to observe the aforementioned breathing motion and to measure the corresponding frequency shift caused by nonlinearities. The possibility of detecting long-time dynamics is strongly constrained by the relatively small propagation distance of the waveguides $(\approx\!10~{\text{cm}})$, a limitation that can nevertheless be overcome through state-recycling techniques~\cite{mukherjee2018state}. Finally, the presence of losses, which would decrease the total guided optical power during its propagation, must be carefully optimized in any experimental realization. 

Cold atoms in optical lattices could offer another versatile platform to observe flat-band phenomena~\cite{Moller2018, Pelegri2019}. Interactions can be tuned in these settings~\cite{bloch2012quantum}, using Feshbach resonances, and disorder (a source of instability) is typically absent. These systems would allow one to investigate deviations from the purely classical regime, and hence, to study the effects of quantum fluctuations on the AB-cages dynamics discussed above. To this aim, we have performed exact-diagonalization calculations, deep in the quantum regime, for the dynamics of $N$ particles ($N \leq 5$) prepared in the initial state $(\hat a^\dag_n)^N|0\ket$, which corresponds to the classical state in Fig.~\ref{fig:lattice}(b) with $\alpha\!=\!0$ when $N\ra \infty$. 
Interestingly, we have found \cite{SuppMat} that leakage from the cage sites, as due to quantum processes, is strongly suppressed already with a small number of particles, and it decreases as $N$ increases. On the other hand, the density inside the cage has a clear pattern and it undergoes a periodic evolution that is reminiscent of collapse and revival dynamics, which manifests in systems with a discrete spectrum \cite{Pitaevskii2016}, such as bosons in a single~\cite{Greiner2002} or double well~\cite{Milburn1997}. These results show that despite leakage (which is detrimental on long time scales), non-trivial ``quasi-caged" quantum dynamics is found over short times deep in the quantum regime. It would therefore be interesting to understand the crossover from the quantum ``quasi-caged" to the classical ``caged" regimes explored in this work.

During the finalization of this manuscript, we became aware of another study dedicated to the impact of nonlinearities in photonic Aharonov-Bohm cages~\cite{Gligoric2018}.


{\it Acknowledgements.$-$} We thank G. Salerno and R. R. Thomson for helpful discussions. N.G.~and M.D.~acknowledge support from the ERC Starting Grant TopoCold. S.M. thanks Universit\'e Libre de Bruxelles (ULB) and Scottish Universities Physics Alliance (SUPA) for hosting and funding through the Postgraduate, Postdoctoral and Early Career Researcher Short-Term Visits Programme-2018, respectively.\\

\bibliography{AB-caging-bib}

\clearpage

\onecolumngrid

\begin{center}
\large
\textbf{Supplemental Material: Nonlinear dynamics of Aharonov-Bohm cages}
\end{center}

\renewcommand{\theequation}{S\arabic{equation}}
\renewcommand{\thefigure}{S\arabic{figure}}
\renewcommand{\thesection}{S\arabic{section}}
\setcounter{figure}{0}
\setcounter{equation}{0}
\setcounter{section}{0}

\twocolumngrid

\begin{figure}[!bp]
\center
\includegraphics[width=0.95\columnwidth]{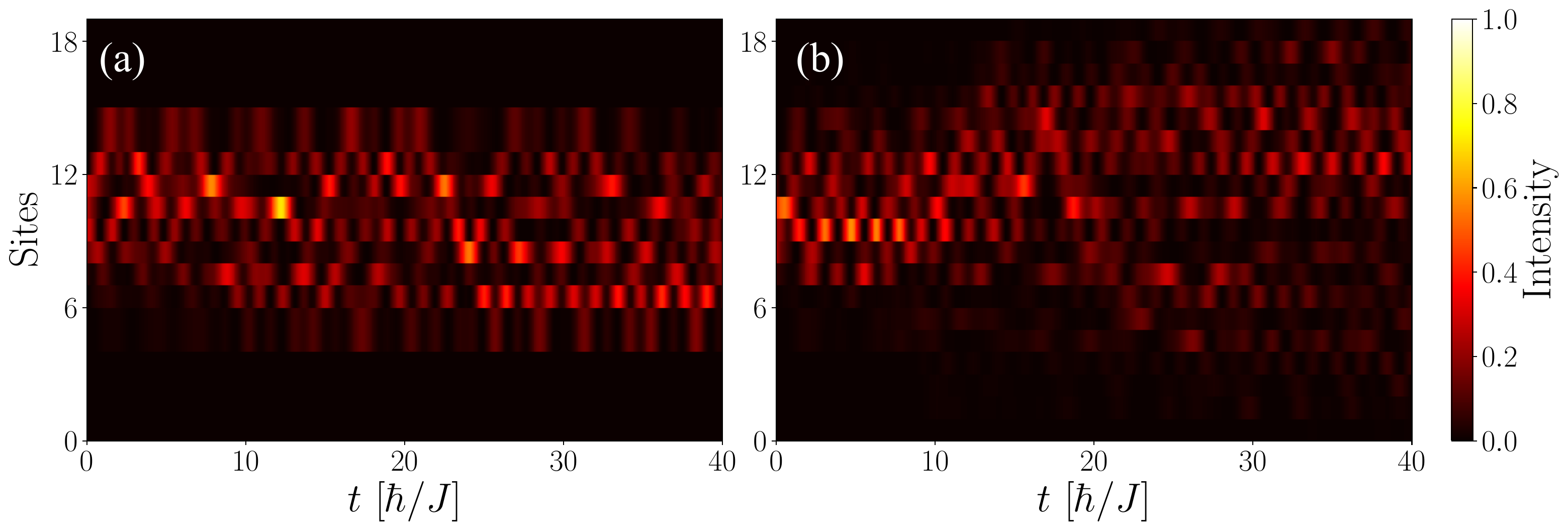}

\caption{(a) Classical time evolution of a random configuration with onsite ($U$) and nearest-neighbor ($V_{NN})$ nonlinearity showing caged dynamics for $U=V_{NN}$ and $g=6$. (b) Same as in (a) but with next-nearest-neighbor nonlinearity $V_{NNN}$ showing loss of caging and propagation of the field over the entire lattice.}
\label{fig:random}
\end{figure}

\section{Dynamics of a random initial state with short- and long-range interactions}

The discrete nonlinear Schr\"odinger equations on the rhombic chain discussed in the main text always provide caged dynamics, where the cage is made of sites that are bounded by two $A_n$ sites, which therefore play the role of a hub. This fact is a consequence of the locality of interactions and can be immediately understood if one considers the two relevant cases: 
\begin{itemize}
\item[1)] if a $A_n$ site has a vanishing amplitude at time $t$ and is one of the boundary sites of the cage, the equations of motion for the sites outside the cage will have no dynamics because they are uncoupled from the cage, thus keeping a zero amplitude over time; 
\item[2)] if a $A_n$ site has a nonvanishing amplitude at time $t$ and there is no amplitude for $m<n$ (or $m>n$), the dynamics of the external $B_{n-1}$ and $C_{n-1}$ sites (or $B_n$ and $C_n$) will be symmetric in amplitude and in-phase (or out-of-phase), as for the state considered in the main text. Therefore, destructive interference will take place on the site $A_{n-1}$ (or $A_{n+1}$), which becomes one of the hubs of the cage as in point 1).
\end{itemize}

\begin{figure}[!bp]
\center
\includegraphics[width=0.95\columnwidth]{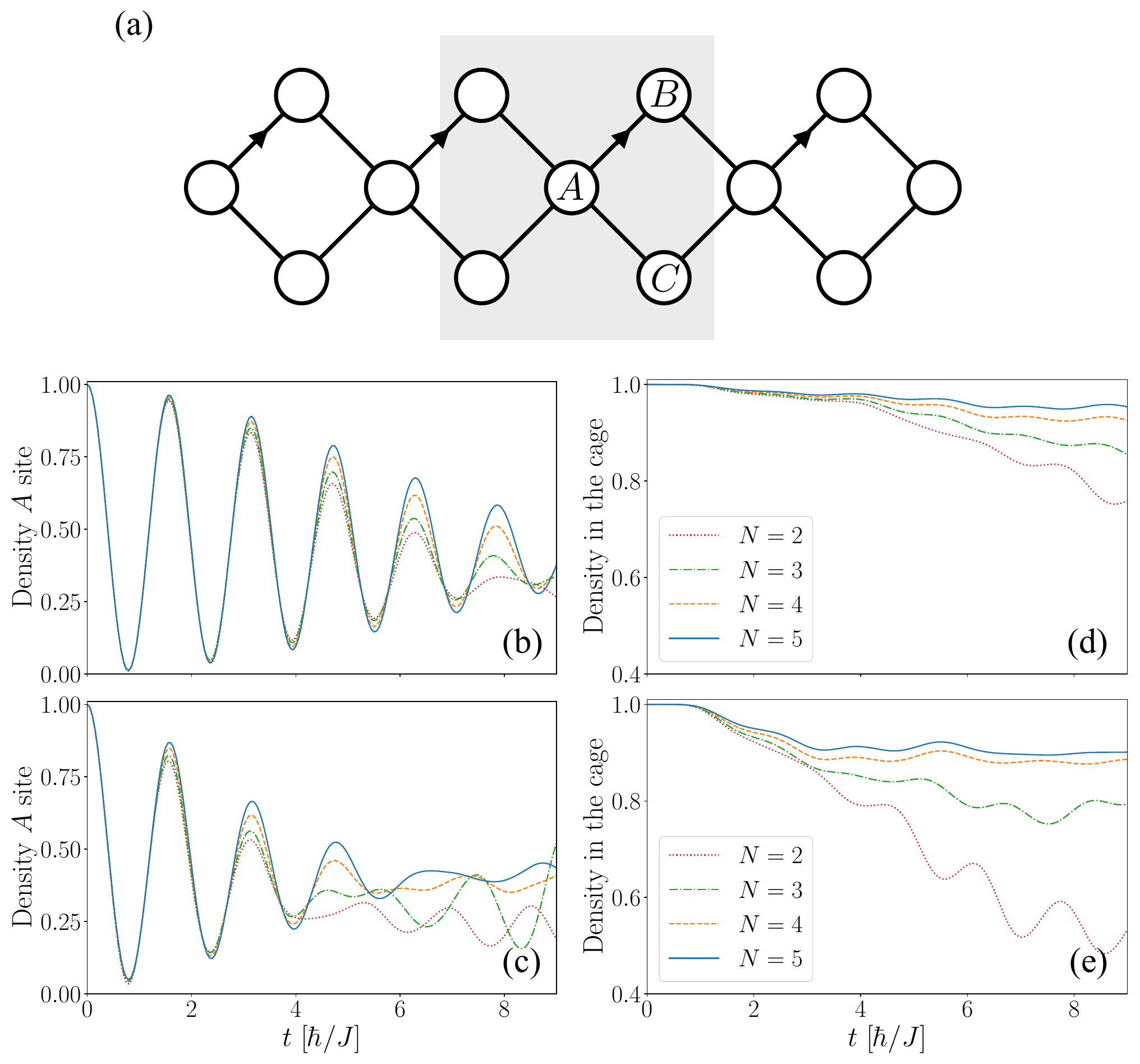}

\caption{(a) Lattice with $L=13$ sites and $\pi$-flux used for the exact-diagonalization calculations in the quantum regime. The arrows indicate the gauge choice used here and in the main text. The shadowed region includes the sites forming the smallest cage in the classical limit.
(b-c) Normalized density time evolution of the central $A$ site for (b) $g=1.2\,J$ and (c) $g=2.4\,J$ with $N=2,3,4,5$ bosons. (d-e) Normalized total density inside the cage sites highlighted in (a) with the same values of $g$ as in (b), (c).}
\label{fig:quantdyn}
\end{figure}

The cases described above build on the locality of interactions, which preserves the interference process that is behind the caging phenomenon. A simple inspection of the equations shows that even nearest-neighbor interactions do not change these conclusions. Indeed, the peripheral $B$ and $C$ sites discussed above in 2) are nevertheless coupled to the same $A_n$ hub site and the symmetry properties of their equations discussed in 2) is not affected. 

Instead, when next-nearest-neighbor interactions are included, $B$ and $C$ sites in neighboring unit cells are directly coupled. Therefore, a density imbalance or a random phase difference between $B$ and $C$ sites in one unit cell will generate an asymmetric dynamics in the neighboring unit cells. The interference process required for caging cannot take place, thus yielding the field to propagate across the entire lattice. 

These conclusions have been tested using a random initial configuration in the case of $i)$ onsite and nearest-neighbor interactions and $ii)$ onsite and next-nearest-neighbor interactions. In Figs.~\ref{fig:random}(a), (b), it is shown that caging is indeed preserved for $i)$ and lost for $ii)$.

\begin{figure}[!tbp]
\center
\includegraphics[width=1\columnwidth]{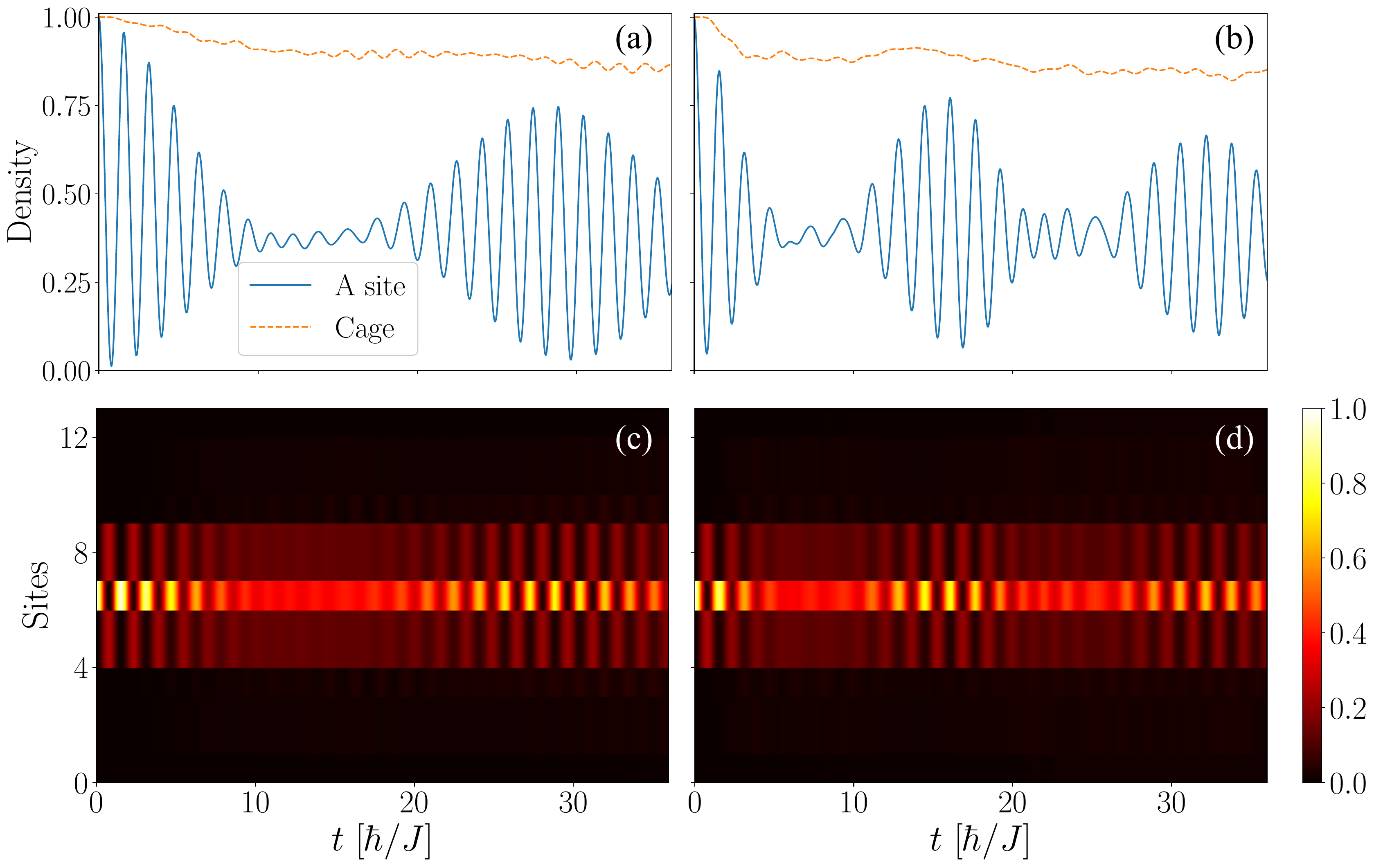}

\caption{(a-b) Normalized density time evolution for $N=4$ bosons and (a) $g=1.2\,J$ and (b) $g=2.4\,J$  showing collapse and revival features (blue line). The dashed red line tracks the total density inside the cage of Fig.~\ref{fig:quantdyn}(a). (c-d) Normalized density time evolution of the full rhombic chain, parameters as in (a), (b).
}
\label{fig:revival}
\end{figure}

\section{Quantum dynamics}

We have performed exact-diagonalization calculations with $N=2,3,4,5$ particles in a rhombic chain with $L=13$ sites (four complete rhombis, see Fig.~\ref{fig:quantdyn}(a)) using a Bose-Hubbard model with onsite interactions described by the Hamiltonian
\be
\hat H = -J \sum_{\mv{i,j}} ( e^{i\theta_{ij} }\hat \psi^\dag_i \hat \psi_j + \textrm{H.c.}) + \f U 2 \sum_i \hat n_i (\hat n_i -1)\,,
\ee
where the Peierls phases $\theta_{ij}$ are chosen such as to have a $\pi$-flux per plaquette and the fields $\hat \psi_i$ describe particles on the $A,B,C$ sites (when $i=3n$, $\hat \psi_{3n} \equiv \hat a_n$, $\hat \psi_{3n+1} \equiv \hat b_n$ and $\hat \psi_{3n+2} \equiv \hat c_n$). We have chosen an initial state with all particles in the central $A$ site 
\be 
|\psi(t=0) \ket = \f{1}{\sqrt{N!}} (\hat a^\dag_n)^N|0\ket\,, 
\ee
and explored several values of the onsite interaction $U$, keeping the product $g = N U$ fixed. In the large $N$ limit, $N\ra\infty$, the initial state corresponds to the classical state with $\alpha=0$ discussed in the main text.

In Figs.~\ref{fig:quantdyn}(b-c), we plot the short time-scale evolution of the density in the central $A$ site for different values of $N$, which shows a clear damped oscillation. In Figs.~\ref{fig:quantdyn}(d-e), we plot the density within the five cage sites of Fig.~\ref{fig:quantdyn}(a) and we observe that the amount of leaked density decreases substantially as we increase $N$. 

The suppression of leaked density raises the question whether quantum caged dynamics with well defined properties already takes place with few particles per site. This is not the case with $N=2$ already at small $g<J$: after few damped oscillations the density completely spreads outside the cage. However, when increasing the number of particles one observes a peculiar evolution inside the cage that shows collapse and revival features (see Fig.~\ref{fig:revival}(a)-(b)), whereas the leaked density remains small over several revival periods.

\end{document}